\documentclass[preprint,onecolumn,10pt,showpacs,aps]{revtex4}%
\usepackage{amsfonts}
\usepackage{amsmath}
\usepackage{amssymb}
\usepackage{graphicx}%
\begin{document}
\title{Implementation of three-qubit Grover search in cavity QED}
\author{W.L. Yang$^{1,2},$ C.Y. Chen$^{3},$ and M. Feng$^{1}$}
\email{mangfeng1968@yahoo.com}
\affiliation{$^{1}$State Key Laboratory of Magnetic Resonance and Atomic and Molecular
Physics, Wuhan Institute of Physics and Mathematics, Chinese Academy of
Sciences, Wuhan 430071, China }
\affiliation{$^{2}$Graduate School of the Chinese Academy of Sciences, Bejing 100049, China}
\affiliation{$^{3}$Department of Physics and Information Engineering, Hunan Institute of
Humanities, Science and Technology, Loudi 417000, China}


\begin{abstract}
Using resonant interaction of three Rydberg atoms with a single-mode microwave
cavity, we consider a realization of three-qubit Grover search algorithm in
the presence of weak cavity decay, based on a previous idea for three-qubit
quantum gate [Phys. Rev. A \textbf{73}, 064304 (2006)]. We simulate the
searching process under the influence of the cavity decay and show that our
scheme could be achieved efficiently to find the marked state with high
fidelity. The required operations are very close to the reach with current
cavity QED techniques.

\end{abstract}
\maketitle

Quantum algorithms, such as Shor's algorithm [1], Grover search [2], and
Deutsch-Jozsa's algorithm [3], have attracted much interest since they could
work in quantum computers, which are in principle able to solve intractable
computational problems more efficiently than present classical computers. Many
efforts have been devoted to achievement of these quantum algorithms
theoretically and experimentally by using trapped ions [4-6], NMR system
[7-8], superconducting mesocircuits [9], cavity quantum electrodynamics (QED)
[10-14], and linear optical elements [15,16].

In this Brief Report, we focus on a scheme of three-qubit Grover search with
cavity QED. Cavity QED has been considering to be an efficient candidate for
small-scale quantum information processing and for quantum network. The rapid
development in relevant experimental technologies has enabled us to achieve
entanglement between two atoms in a microwave cavity [11], based on which
\ there have been some proposals for two-qubit Grover search with cavity QED
[12,13]. We have also noticed a very recent publication for three-qubit Grover
search with three four-level atoms going through a three-mode cavity [14].
Actually, the important difference of the three-qubit Grover search from the
two-qubit case is the probabilistic achievement. To reach a case with high
success probability, we have to implement the basic searching step (also
called iteration) for several times. So implementation of a three-qubit Grover
search is much more complex than that of a two-qubit case. In contrast to
[14], we will design a simpler but efficient Grover search scheme by three
identical Rydberg atoms sent through a single-mode microwave cavity. We will
store quantum information in long-lived internal levels of the Rydberg atoms,
and consider the resonant interaction between the atoms and the cavity mode,
which yields a very fast implementation of the search. As the cavity decay is
the main dissipative factor of our design, we will seriously consider its
detrimental effect on our scheme.

Let us first briefly review the main points of a Grover search algorithm,
which consists of three kinds of operations [4]. The first one is to prepare a
superposition state $\left\vert \Psi_{0}\right\rangle =(\frac{1}{\sqrt{N}%
})\sum_{i=0}^{N-1}\left\vert i\right\rangle $ using Hadamard gates. The second
is for an iteration $Q$ including following two operations: (a) Inverting the
amplitude of the marked state $\left\vert \tau\right\rangle $ using a quantum
phase gate $I_{\tau}=I-2\left\vert \tau\right\rangle \left\langle
\tau\right\vert ,$ with $I$ the identity matrix; \ (b) Inversion about average
of the amplitudes of all states using the diffusion transform $\allowbreak
\hat{D}$, with $\hat{D}_{ij}=\tfrac{2}{N}-\delta_{ij}\ $($i,j=1,2,3,$%
\textperiodcentered\textperiodcentered\textperiodcentered\textperiodcentered
\textperiodcentered\textperiodcentered$\allowbreak N$) and $\allowbreak
N=2^{q}$ ($q$ $\allowbreak$being the qubit number). This step should be
carried out by at least $\pi\sqrt{N}/4$ times to maximize the probability for
finding the marked state. Finally, a measurment of the whole system is done to
get the marked state. In other word, the Grover search consists in a
repetition of the transformation $\allowbreak Q=HI_{000}HI_{\tau}$ with
$I_{000}=I-2\left\vert 000\right\rangle \left\langle 000\right\vert $ (defined later).

In three-qubit case, the number of possible quantum states is $\allowbreak
2^{3}$, and the operation to label a marked state by conditional quantum phase
gate is $\allowbreak I_{\tau}$ with $\tau$ one of the states $\{\left\vert
000\right\rangle ,\left\vert 001\right\rangle ,\left\vert 010\right\rangle
,\left\vert 011\right\rangle ,\left\vert 100\right\rangle ,\left\vert
101\right\rangle ,\left\vert 110\right\rangle ,\left\vert 111\right\rangle
\}.$ For clarity of description, we first consider an ideal situation. For
three identical atoms, the atomic internal states under our consideration are
denoted by $\left\vert i_{j}\right\rangle ,\left\vert g_{j}\right\rangle ,$and
$\left\vert e_{j}\right\rangle ,$ with $\left\vert g_{j}\right\rangle $ and
$\left\vert i_{j}\right\rangle $ being states lower than $\left\vert
e_{j}\right\rangle $. Because the resonant transition happens between
$\left\vert g_{j}\right\rangle $\ and $\left\vert e_{j}\right\rangle $ by the
cavity mode, $\left\vert i_{j}\right\rangle $ is not involved in the
interaction throughout our scheme. So the Hamiltonian in units of $\hbar=1$
reads,$\ \ \ \ \ \ \ \ \ \ \ $%
\begin{equation}
\ \ \ \allowbreak H=\sum\limits_{j=1}^{3}\Omega_{jc}(a^{+}S_{j}^{-}+aS_{j}%
^{+}), \label{1}%
\end{equation}
where $\Omega_{jc}$\ is the coupling constant of the $\allowbreak j$th atom to
the cavity mode, $S_{j}^{+}=\left\vert e_{j}\right\rangle \left\langle
g_{j}\right\vert $ and $S_{j}^{-}=\left\vert g_{j}\right\rangle \left\langle
e_{j}\right\vert $\ are the atomic spin operators for raising and lowering,
respectively, and $a^{+}$\ $(\allowbreak a)$\ is the creation (annihilation)
operator for the cavity mode. Following the proposal by sending atoms through
a microwave cavity simultaneously [17], to achieve three-qubit conditional
phase gate, we require that the three atoms couple to the cavity mode by
$\Omega_{1c}:\Omega_{2c}:\Omega_{3c}=1:\sqrt{35}:8$ and the gating time be
$\dfrac{\pi}{\Omega_{1c}}.$\ In our proposal, the qubit definitions are not
the same for each atom. The logic state $\left\vert 1\right\rangle $
($\left\vert 0\right\rangle $) of the qubit 1 is denoted by $\left\vert
g_{1}\right\rangle $ ($\left\vert e_{1}\right\rangle )$ of the atom 1;
$\left\vert g_{2}\right\rangle $ and $\left\vert i_{2}\right\rangle $\ of\ the
atom 2 encode the logic state $\left\vert 1\right\rangle $ ($\left\vert
0\right\rangle $) of the qubit 2; The logic state $\left\vert 1\right\rangle $
($\left\vert 0\right\rangle $) of the qubit 3 is represented by $\left\vert
g_{3}\right\rangle $ ($\left\vert i_{3}\right\rangle )$ of the atom 3. Ref.
[17] has shown us the possibility to achieve an approximate three-qubit
quantum phase gate $\allowbreak I_{e_{1}i_{2}i_{3}}=\allowbreak I_{000}=diag$
$\{-1,\gamma_{0},1,1,1,1,1,1\}$ in a computational subspace spanned by
$\left\vert e_{1}\right\rangle \left\vert i_{2}\right\rangle \left\vert
i_{3}\right\rangle ,$ $\left\vert e_{1}\right\rangle \left\vert i_{2}%
\right\rangle \left\vert g_{3}\right\rangle ,\left\vert e_{1}\right\rangle
\left\vert g_{2}\right\rangle \left\vert i_{3}\right\rangle ,\left\vert
e_{1}\right\rangle \left\vert g_{2}\right\rangle \left\vert g_{3}\right\rangle
,$ $\left\vert g_{1}\right\rangle \left\vert i_{2}\right\rangle \left\vert
i_{3}\right\rangle ,$ $\left\vert g_{1}\right\rangle \left\vert i_{2}%
\right\rangle \left\vert g_{3}\right\rangle ,$ $\left\vert g_{1}\right\rangle
\left\vert g_{2}\right\rangle \left\vert i_{3}\right\rangle ,$ $\left\vert
g_{1}\right\rangle \left\vert g_{2}\right\rangle \left\vert g_{3}\right\rangle
,$ where $\gamma_{0}=\frac{\Omega_{1c}^{2}}{\Omega_{1c}^{2}+\Omega_{3c}^{2}%
}\cos(\sqrt{65}\pi)+$ $\frac{\Omega_{3c}^{2}}{\Omega_{1c}^{2}+\Omega_{3c}^{2}%
}=0.9997.$ To carry out the Grover search, we define the three-qubit Hadamard
gate,%
\begin{equation}
H^{\otimes3}=\prod\limits_{i=1}^{3}H_{i}=\left(  \frac{1}{\sqrt{2}}\right)
^{3}%
\begin{bmatrix}
1 & 1\\
1 & -1
\end{bmatrix}
\otimes%
\begin{bmatrix}
1 & 1\\
1 & -1
\end{bmatrix}
\otimes%
\begin{bmatrix}
1 & 1\\
1 & -1
\end{bmatrix}
,
\end{equation}
where $H_{i}$ is the Hadamard gate acting on the $\allowbreak i$th atom,
transforming states as $\left\vert e_{1}\right\rangle \rightarrow(1/\sqrt
{2})(\left\vert e_{1}\right\rangle +\left\vert g_{1}\right\rangle ),$
$\left\vert g_{1}\right\rangle \rightarrow(1/\sqrt{2})(\left\vert
e_{1}\right\rangle -\left\vert g_{1}\right\rangle ),$ $\left\vert
i_{2}\right\rangle \rightarrow(1/\sqrt{2})(\left\vert i_{2}\right\rangle
+\left\vert g_{2}\right\rangle ),$ $\left\vert g_{2}\right\rangle
\rightarrow(1/\sqrt{2})(\left\vert i_{2}\right\rangle -\left\vert
g_{2}\right\rangle ),$ $\left\vert i_{3}\right\rangle \rightarrow(1/\sqrt
{2})(\left\vert i_{3}\right\rangle +\left\vert g_{3}\right\rangle ),$
$\left\vert g_{3}\right\rangle \rightarrow(1/\sqrt{2})(\left\vert
i_{3}\right\rangle -\left\vert g_{3}\right\rangle ).$ These gatings could be
performed by external microwave pulses.

It is easy to find that the transformation $\allowbreak Q=\ \allowbreak
H^{\otimes3}I_{000}\ \allowbreak H^{\otimes3}I_{\tau}=\allowbreak H^{\otimes
3}\allowbreak I_{e_{1}i_{2}i_{3}}\ \allowbreak H^{\otimes3}I_{\tau}=-\hat
{D}I_{\tau},$ which implies that the diffusion transform $\hat{D}$ is always
unchanged, no matter which state is to be searched. The only change is the
phase gate $I_{\tau}$ for different marked states. Based on the gate $I_{000}$
to mark the state $\left\vert e_{1}i_{2}i_{3}\right\rangle ,$ we could
construct other seven gates for the marking job as,%
\begin{align}
\ \ \ I_{e_{1}i_{2}g_{3}}  &  =I_{001}=\sigma_{x,3}I_{000}\sigma
_{x,3},\ \ I_{e_{1}g_{2}i_{3}}=I_{010}=\sigma_{x,2}I_{000}\sigma_{x,2},\text{
}I_{e_{1}g_{2}g_{3}}=I_{011}=\sigma_{x,3}\sigma_{x,2}I_{000}\sigma_{x,2}%
\sigma_{x,3},\nonumber\\
I_{g_{1}i_{2}i_{3}}  &  =I_{100}=\sigma_{x,1}I_{000}\sigma_{x,1}%
,\ \ \ I_{g_{1}i_{2}g_{3}}=I_{101}=\sigma_{x,3}\sigma_{x,1}I_{000}\sigma
_{x,1}\sigma_{x,3},\text{ }I_{g_{1}g_{2}i_{3}}=I_{110}=\sigma_{x,2}%
\sigma_{x,1}I_{000}\sigma_{x,1}\sigma_{x,2},\nonumber\\
I_{g_{1}g_{2}g_{3}}  &  =I_{111}=\sigma_{x,3}\sigma_{x,2}\sigma_{x,1}%
I_{000}\sigma_{x,1}\sigma_{x,2}\sigma_{x,3}.
\end{align}
So with the state marked, and the three-qubit diffusion transform $\hat{D}%
$\ which is generated by combining two Hadamard gates $H^{\otimes3}$ with the
quantum phase gate $I_{000}$, a full Grover search for three qubits is available.

Taking the marked state $\left\vert 101\right\rangle $ as an example, we
design a three-qubit Grover search setup in Fig. 1. The cavity is a microwave
cavity sustaining a single mode with a standing-wave pattern along the z-axis.
The atoms 1, 2 and 3 prepared in high-lying circular Rydberg states are sent
through the cavity with proper speed, resonantly interacting with the cavity
mode. Single-qubit rotations are made at certain times by external microwave
pulses, and the state-selective field-ionization detectors $\allowbreak D_{1}%
$, $D_{2}$, $\allowbreak D_{3}$ are settled at the end of the passage for
checking the states of the atoms 1, 2 and 3, respectively. One point to
mention is that, in searching the state $\left\vert 011\right\rangle $ or
$\left\vert 111\right\rangle ,$ imhomogeneous electric fields are needed to
tune the atomic transitions through the Stark effect [12,13], which make the
single-qubit operations completed individually. But these imhomogeneous
electric fields are unnecessary in searching other states.

As the resonant interaction actually excites the cavity mode, although we
could carry out the scheme very fast, we should consider the cavity decay
seriously. Under the assumption of weak cavity decay that no photon actually
leaks out of the microwave cavity during our implementation time, we employ
the quantum trajectory method by the Hamiltonian,\
\begin{equation}
\ \ \ \ \allowbreak H=\sum\limits_{j=1}^{3}\Omega_{j_{c}}(a^{+}S_{j}%
^{-}+aS_{j}^{+})-i\frac{\kappa}{2}a^{+}a,
\end{equation}
where $\kappa$ is the cavity decay rate. As discussed in [17], under the weak
decay condition, the cavity dissipation only affects the diagonal elements of
the matrix for the phase gate. For example, by choosing the interaction time
$\allowbreak t_{I}=\pi/A_{1\kappa}$ with $A_{1\kappa}=\sqrt{\Omega_{1c}%
^{2}-\kappa^{2}/16}$ and the condition $\Omega_{1c}:\Omega_{2c}:\Omega
_{3c}=1:\sqrt{35}:8$, we generate the three-qubit phase gate $I_{000}%
^{^{\prime}}$\ in the decay case,\ $\allowbreak$%
\begin{equation}
I_{e_{1}i_{2}i_{3}}^{^{\prime}}=diag\{-\mu_{1},\gamma_{1},\beta_{1},\alpha
_{1},1,1,1,1\}=U_{0}(t),
\end{equation}
where $\alpha_{1}=1-\frac{\Omega_{1c}^{2}}{\Omega_{1c}^{2}+\Omega_{2c}%
^{2}+\Omega_{3c}^{2}}(1-e^{-\kappa t/4}),$ \ $\beta_{1}=1-\frac{\Omega
_{1c}^{2}}{\Omega_{1c}^{2}+\Omega_{2c}^{2}}(1-e^{-\kappa t/4}),$ $\mu
_{1}=e^{-\kappa t/4},$ and $\gamma_{1}=1-\frac{\Omega_{1c}^{2}}{\Omega
_{1c}^{2}+\Omega_{3c}^{2}}[1$ $-$ $e^{-\kappa t/4}\cos(\sqrt{65}\pi)]$ after
the negligible term $\frac{\kappa}{4A_{1\kappa}}\sin(A_{1\kappa})$ is omitted.
So for a state $\left\vert \Psi\right\rangle =\frac{1}{2\sqrt{2}}(\bar{A}%
_{j}\left\vert e_{1}i_{2}i_{3}\right\rangle +B_{j}\left\vert e_{1}i_{2}%
g_{3}\right\rangle +$ $\allowbreak C_{j}\left\vert e_{1}g_{2}i_{3}%
\right\rangle $ $+D_{j}\left\vert e_{1}g_{2}g_{3}\right\rangle +E_{j}%
|g_{1}i_{2}i_{3}\rangle+F_{j}\left\vert g_{1}i_{2}g_{3}\right\rangle
+G_{j}\left\vert g_{1}g_{2}i_{3}\right\rangle +H_{j}\left\vert g_{1}g_{2}%
g_{3}\right\rangle ),$ the success probability of the phase gate is defined as%
\begin{equation}
P_{j}=(|D_{j}|^{2}\alpha_{i}^{2}+|C_{j}|^{2}\beta_{i}^{2}+|B_{j}|^{2}%
\gamma_{i}^{2}+|\bar{A}_{j}|^{2}\mu_{i}^{2}+|E_{j}|^{2}+|F_{j}|^{2}%
+|G_{j}|^{2}+|H_{j}|^{2})/8,
\end{equation}
where $\allowbreak j=0,1$ correspond to the ideal and decay cases,
respectively, with $\alpha_{0}=\beta_{0}=\mu_{0}=1$. In our case, the atomic
system is initially prepared in $\left\vert \Psi_{0}\right\rangle =\frac
{1}{2\sqrt{2}}(\left\vert g_{1}\right\rangle +\left\vert e_{1}\right\rangle
)(\left\vert g_{2}\right\rangle +\left\vert i_{2}\right\rangle )(\left\vert
g_{3}\right\rangle +\left\vert i_{3}\right\rangle ),$ which corresponds to a
success probability of the three-qubit phase gate $P_{j}=(4+\alpha_{j}%
^{2}+\beta_{j}^{2}+\gamma_{j}^{2}+\mu_{j}^{2})/8.$

As mentioned previously, the three-qubit Grover search is carried out only
probabilistically. So how to obtain a high success rate of the search is the
problem of much interest, particularly in the presence of weak cavity decay.
We have numerically simulated the Grover search for finding different marked
states in the cases of $\kappa=0$\ (the ideal case), $\kappa=\Omega_{1c}/50$,
and $\kappa=\Omega_{1c}/10.$ Due to the similarity, we only demonstrate the
search for a marked state $\left\vert e_{1}\right\rangle \left\vert
i_{2}\right\rangle \left\vert i_{3}\right\rangle $ in Fig. 2 as an example.
Considering the success rates of the three-qubit phase gating (i.e., Eq. (6))
and the Grover search itself, we show in Fig. 2(a) that the success
probability is smaller and smaller with the increase of the decay rate and the
iteration number. This implies that, although the sixth iteration could reach
the largest success rate in the ideal consideration, we prefer the second
iteration in the presence of dissipation. The detrimental effect from the
cavity decay is also reflected in the estimate of fidelity in Fig. 2(b).

We briefly address the experimental feasibility of our scheme with current
microwave cavity technology by considering three Rydberg atoms with principal
quantum numbers 49, 50 and 51 to be levels $\left\vert i\right\rangle $,
$\left\vert g\right\rangle $ and $\left\vert e\right\rangle ,$ respectively.
Based on the experimental numbers reported in [10,11], the coupling strength
at the cavity centre could be $\Omega_{0}=2\pi\times49kHz$, and the Rydberg
atomic lifetime is 30 ms. Since the single-qubit operation takes negligible
time in comparison with that for the three-qubit phase gating, an iteration of
our proposed Grover search would take $\allowbreak t_{0}=2\pi/\sqrt
{\Omega_{1c}^{2}-\kappa^{2}/16}$. Direct calculation shows that the time for
one iteration is about$\ 160\mu s$, much shorter than the cavity decay time
for both cases of $\kappa=\Omega_{1c}/50$, and $\Omega_{1c}/10.$ So our
treatment with quantum trajectory method is physically reasonable.

With current cavity QED techniques, the design in Fig. 1 should be realized by
four separate microwave cavities with each Ramsey zone located by a cavity.
Since each microwave cavity is employed to carry out a three-qubit phase gate
I$_{000},$ the four cavities should be identical. While for searching
different states, we employ different single-qubit operations, as shown in Eq.
(3). So the Ramsey zones should be long enough to finish at most two
consecutive single-qubit operations, for example, to search states $\left\vert
g_{1}g_{2}g_{3}\right\rangle $, we have to carry out a Hadamad gate H and a
gate I$_{111}$ including three simultaneous single-qubit rotations. Above
requirements are due to the fact that each atom is sent by a fix velocity to
fly through the design in Fig. 1, and each single-qubit operation takes a time
(although it is very short so that we roughly omitted this time in above
assessment of the implementation time). In principle, if each component of the
design is available, our scheme would be achievable experimentally. However,
we have not yet found an experimental report for three atoms simultaneously
going through a microwave cavity, and the two-atom entanglement in a microwave
cavity was done by using van der Waals collision between the atoms [11] under
a non-resonant condition. Nevertheless, compared to [14] with four-level atoms
sent through a three-mode cavity, our proposal involving a single-mode cavity
is much simpler and is closer to the reach with the current cavity QED
technology. Considering the intra-atom interaction occurs in the central
region of the cavity, we have $\allowbreak\Omega_{jc}\simeq\Omega_{0}\cos(2\pi
z/\lambda_{0}).$ So the three atoms should be sent through the cavity with the
atom 3 going along the $\allowbreak y$-axis ($\allowbreak x_{3}$=$\allowbreak
z_{3}$=0) and atoms 1 and 2 away from the atom 3\ by $\left\vert
\allowbreak\allowbreak z_{1}\right\vert /\left\vert \allowbreak\allowbreak
z_{2}\right\vert =$ $\arccos(0.125)/\arccos(\sqrt{35}/8)$ $\approx1.957.$ With
the manipulation designed in Fig. 1, a three-qubit Grover search for the
marked state $\left\vert 101\right\rangle $ would be achievable.

We have noticed that four-qubit Grover search with linear optical elements has
been achieved [16]. While as photons are always flying, they are actually
unsuitable for a practical quantum computing. In contrast, the atoms move much
more slowly than photons, and are thereby relatively easier for manipulation.
In addition, the three-qubit gating we employed simplifies the implementation
and reduces the probability of error in comparison with the series of
two-qubit gatings in [16]. More importantly, our scheme could be
straightforwardly applied to the ion-trap-cavity combinatory setup [18] or
cavity-embedded optical lattices confining atoms [19], in which the atoms are
localized and the model we employ here still works. For these considerations,
however, the cavities should be optical ones, for which we have to consider
both the cavity decay and the atomic spontaneous emission. Based on a previous
treatment [20], as long as these dissipations are weak, the three-qubit phase
gating would also be available, and thereby our scheme is in principle
workable in optical regime.

Besides the imperfection considered above, there are other unpredictable
imperfection in an actual experiment, such as diversity in atomic velocities,
deflected atomic trajectories, classical pulse imperfection, slight difference
of the cavities and so on. Let us take two examples to assess the influence
from imperfection$.$ First, as it is still a challenge to simultaneously send
three Rydberg atoms through a cavity with precise velocities in experimental
performance, we consider an imperfection in this respect. For the clarity and
convenience of our discusssion, we simply consider a situation that the atom 1
moves a little bit slower than the atoms 2 and 3, i.e., the times of the atoms
passing through the cavity $t_{1}=t_{0}+\delta t$ and $t_{2}=t_{3}=t_{0},$
with $\allowbreak t_{0}$ the desired interaction time for the three-qubit
phase gate $I_{000}^{^{\prime}}.$ Direct calculation yields the infidelity in
a single three-qubit phase gate due to the imperfection in atomic velocity to
be,
\begin{equation}
Infidelity=1-\frac{[4+\xi\alpha_{1}+\xi\beta_{1}+\xi\mu_{1}+\xi\gamma
_{1}-\Omega_{1c}^{2}/(A_{1\kappa}A_{3\kappa})\exp(-\kappa\delta t/4)\sin
(A_{1\kappa}\delta t)\sin(\sqrt{65}\pi)]^{2}}{8[4+(\xi\alpha_{1})^{2}%
+(\xi\beta_{1})^{2}+(\xi\mu_{1})^{2}+(\xi\gamma_{1}-\Omega_{1c}^{2}%
/(A_{1\kappa}A_{3\kappa})\exp(-\kappa\delta t/4)\sin(A_{1\kappa}\delta
t)\sin(\sqrt{65}\pi))^{2}]},
\end{equation}
where $\xi=\exp(-\kappa\delta t/4)[\cos(A_{1\kappa}\delta t)+\frac{\kappa
}{4A_{1\kappa}}\sin(A_{1\kappa}\delta t)],$ and $A_{3\kappa}=2\pi/\sqrt
{\Omega_{1c}^{2}+\Omega_{3c}^{2}-\kappa^{2}/16}.$ Due to the additional
interaction regarding the atom 1, an enlarging infidelity occurs with respect
to the time difference $\delta t$ and the decay rate $\kappa$, as shown in
Fig. 3. Secondly, we consider the unfavorable influence from the coupling
strength $\Omega_{jc}^{\prime}$ in some cavities with the offset $\eta
\Omega_{jc}$ from the ideal number$.$ We find the infidelity due to these
offsets for a Grover search implementation to be,%
\begin{equation}
Infidelity^{^{\prime}}=1-\frac{(4+\alpha_{\chi}^{^{\prime}}+\beta_{\chi
}^{^{\prime}}+\gamma_{\chi}^{^{\prime}}+\mu_{\chi}^{^{\prime}})^{2}%
}{8(4+\alpha_{\chi}^{^{\prime}2}+\beta_{\chi}^{^{\prime}2}+\gamma_{\chi
}^{^{\prime}2}+\mu_{\chi}^{^{\prime}2})},
\end{equation}
where $\alpha_{\chi}^{^{\prime}}=[1-\frac{\Omega_{1c}^{^{\prime}2}}%
{\Omega_{1c}^{^{\prime}2}+\Omega_{2c}^{^{\prime}2}+\Omega_{3c}^{^{\prime}2}%
}(1-e^{-\kappa t_{0}/4})]^{\chi}\alpha_{1}^{4-\chi},$ $\beta_{\chi}^{^{\prime
}}=[1-\frac{\Omega_{1c}^{^{^{\prime}}2}}{\Omega_{1c}^{^{\prime}2}+\Omega
_{2c}^{^{\prime}2}}(1-e^{-\kappa t_{0}/4})]^{\chi}\beta_{1}^{4-\chi},$
$\gamma_{\chi}^{^{\prime}}=\{1-\frac{\Omega_{1c}^{^{\prime}2}}{\Omega
_{1c}^{^{\prime}2}+\Omega_{3c}^{^{\prime}2}}[1-e^{-\kappa t_{0}/4}\cos
(\sqrt{65}\pi)]\}^{\chi}\gamma_{1}^{4-\chi},$ and $\mu_{\chi}^{^{\prime}%
}=e^{-\chi\kappa t_{0}/4}\mu_{1}^{4-\chi},$ with $\chi$ $(=1,2,3,4)$ the
number of the cavities with the coupling strength offsets. We plot the
dependence of the infidelity on different $\eta$ and $\chi$ in the case of
$\kappa=\Omega_{1c}/10$ in Fig. 4. The error assessments in Figs. 3 and 4 are
actually for the simplest consideration about imperfection. In a realistic
experiment, situation would be more complicated. So to carry out our scheme
efficiently and with high fidelity, we have to suppress these imperfect
factors as much as we can.

In conclusion, we have proposed a potentially practical scheme for realizing a
three-qubit Grover search by resonant interaction of three Rydberg atoms in a
microwave cavity. We have estimated the influence from the cavity decay on our
scheme and shown that large enough success rate and fidelity could be reached
for a three-qubit Grover search with current or near-future technique of
cavity QED. Although we have not yet found our idea to be extendable to more
than three-qubit case, our scheme could be extended to trapped ions embedded
in a cavity or atoms in cavity-embedded optical lattices. So we argue that our
present scheme is helpful for demonstration of Grover search algorithm by
small-scale quantum information processing devices.

{\large ACKNOWLEDGMENTS}

This work is partly supported by NNSF of China under Grant No. 10474118, by
Hubei Provincial Funding for Distinguished Young Scholars, and partly by the
NFRP of China under Grants No. 2005CB724502 and No. 2006CB921203.

Note added: After finishing this work, we became aware of a work for N-qubit
Toffoli gate in a cavity by resonant interaction [21], in which the only
difference from [17] is the different setting of atom-cavity coupling
strength. This means that our idea for Grover search would be in principle
extended to N-qubit case after slight modification.

\bigskip

[1] A.R. Calderbank and P.W. Shor, Phys. Rev. A. \textbf{54}, 1098 (1996).

[2] L.K. Grover, Phys. Rev. Lett. \textbf{79}, 325 (1997); \textit{ibid.}
\textbf{80}, 4329 (1998).

[3] D. Deutsch and R. Jozsa, Proc. R. Soc. London, ser. A. \textbf{439}, 553 (1992).

[4] M. Feng, Phys. Rev. A \textbf{63}, 052308 (2001).

[5] S. Fujiwara and S Hasegawa, Phys. Rev. A. \textbf{71}, 012337 (2005).

[6] K.-A. Brickman \textit{et al}, Phys. Rev. A. \textbf{72}, R050306 (2005).

[7] J.A. Jones, M. Mosca and R.H. Hansen, Nature (London) \textbf{393}, 344(1998).

[8] I.L. Chuang, N. Gershenfeld and M. Kubinec, Phys. Rev. Lett. \textbf{80},
3408 (1998).

[9] Y. Nakamura, Yu. A. Pashkin and J.S. Tsai, Nature (London) \textbf{398},
786 (1999); D. Vio \textit{et al}, Science \textbf{296}, 886 (2002); M.S.
Anwar \textit{et al}, Chem. Phys. Lett. \textbf{400}, 94 (2004).

[10] A. Rauschenbeutel \textit{et al}, Phys. Rev. Lett. \textbf{83}, 5166 (1999).

[11] S. Osnaghi \textit{et al}, Phys. Rev. Lett. \textbf{87} 037902 (2001).

[12] F. Yamaguchi \textit{et al}, Phys. Rev. A. \textbf{66}, R010302 (2002).

[13] Z.J. Deng, M. Feng and K.L. Gao, Phys. Rev. A. \textbf{72}, 034306 (2005).

[14] A. Joshi and M. Xiao, Phys. Rev. A. \textbf{74}, 052318 (2006).

[15] P.G. Kwiat \textit{et al}, J. Mod. Opt. \textbf{47}, 257 (2000).

[16] P. Walther \textit{et al}, Nature (London) \textbf{434}, 169 (2005).

[17] C.Y. Chen, M. Feng and K.L. Gao, Phys. Rev. A \textbf{73}, 064304 (2006).

[18] M. Feng and X. Wang, J. Opt. B: Quantum Semiclass. Opt., \textbf{4}, 283 (2002).

[19] J.A. Sauer \textit{et al}, Phys. Rev. A. \textbf{69}, R051804 (2004).

[20] C.Y. Chen, M. Feng and K.L. Gao, J. Phys. A \textbf{39}, 11861 (2006).

[21] X.Q. Shao \textit{et al},\ Phys. Rev. A \textbf{75}, 034307 (2007).

\textbf{Captions of Figures}

FIG. 1. Schematic setup for finding the marked state $\left\vert
101\right\rangle $ in a three-qubit Grover search. Three atoms initially
prepared in a superposition state $\left\vert \Psi_{0}\right\rangle $\ go
through the cavity with the identical velocity from the box $\allowbreak B$.
We send the atom 3 through the center of the microwave cavity along the y axis
and other two atoms away from the y axis. We consider twice searching steps in
the setup, which yields the largest success rate in the presence of
dissipation. The operations $H^{\otimes3},$ $\sigma_{x,1}$, $\sigma_{x,3}$ and
$U_{0}(t)$ are defined in the text. Only in the case that the marked state is
$\left\vert 011\right\rangle $ or $\left\vert 111\right\rangle ,$ should
additional imhomogeneous electric fields be applied on the regions for
single-qubit operation to distinguish the atoms 2 and 3.

FIG. 2. Numerical results for a three-qubit Grover search for the marked state
$\left\vert e_{1}\right\rangle \left\vert i_{2}\right\rangle \left\vert
i_{3}\right\rangle ,$ where $k=\kappa$ and $g=\Omega_{1c}.$ (a) Probability
for finding the marked state in the case of $\kappa=0,$ $\Omega_{1c}/50$ and
$\Omega_{1c}/10$; (b) Fidelity of the searched state in the case of
$\kappa=\Omega_{1c}/50$ and $\Omega_{1c}/10.$

FIG. 3. Infidelity in a three-qubit phase gate versus time delay$,$ where the
solid and dashed\ curves represent the cases of $\kappa=\Omega_{1c}/50$ and
$\Omega_{1c}/10,$ respectively.

FIG. 4. Infidelity in a Grover search versus offset constant $\eta,$ where the
four solid curves from bottom to top correspond to the number of imperfect
cavities varying from 1 to 4 in the case of $\kappa=\Omega_{1c}/10$.

\newpage

\newpage%
\begin{figure}
[ptb]
\begin{center}
\includegraphics[
height=11.361in,
width=7.8923in
]%
{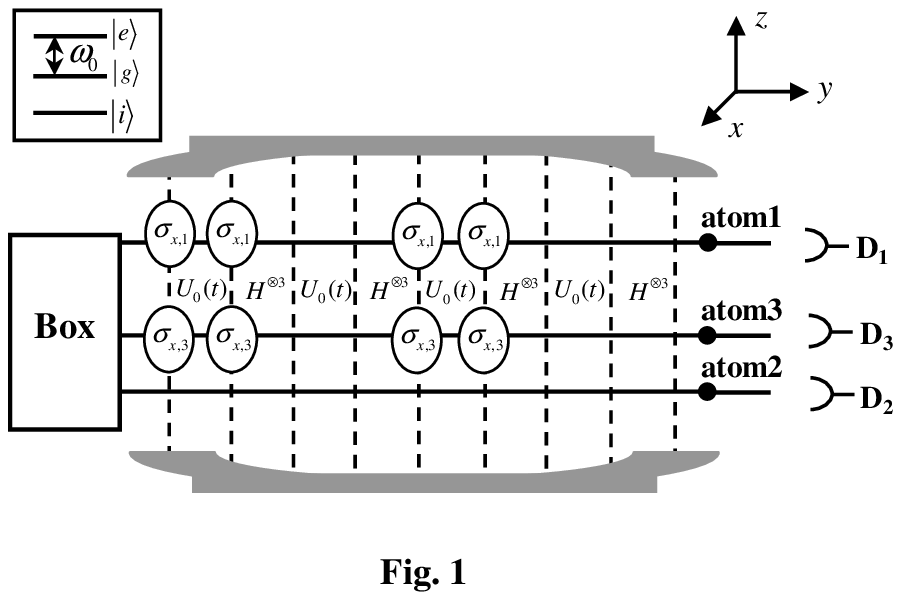}%
\end{center}
\end{figure}

\newpage

\newpage%
\begin{figure}
[ptb]
\begin{center}
\includegraphics[
height=11.361in,
width=7.8923in
]%
{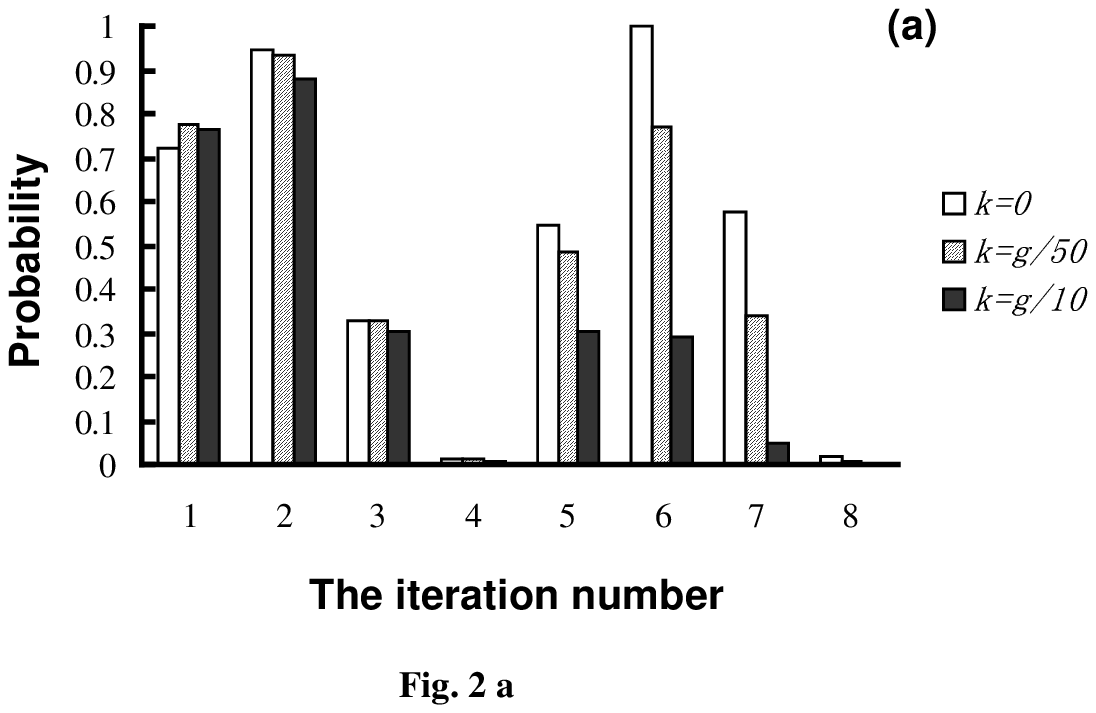}%
\end{center}
\end{figure}
\begin{figure}
[ptbptb]
\begin{center}
\includegraphics[
height=11.361in,
width=7.8923in
]%
{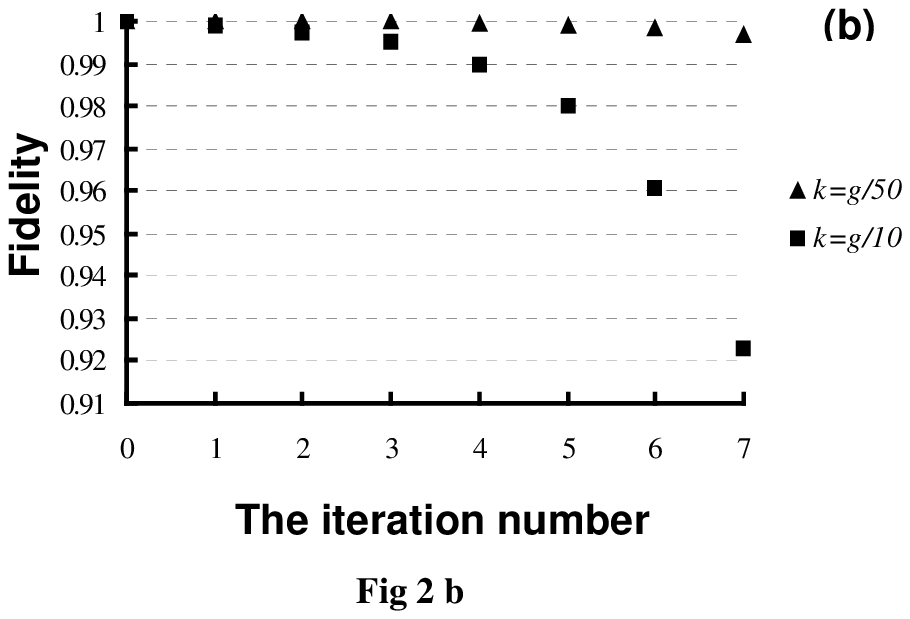}%
\end{center}
\end{figure}

\newpage

\newpage%
\begin{figure}
[ptb]
\begin{center}
\includegraphics[
height=11.361in,
width=7.8923in
]%
{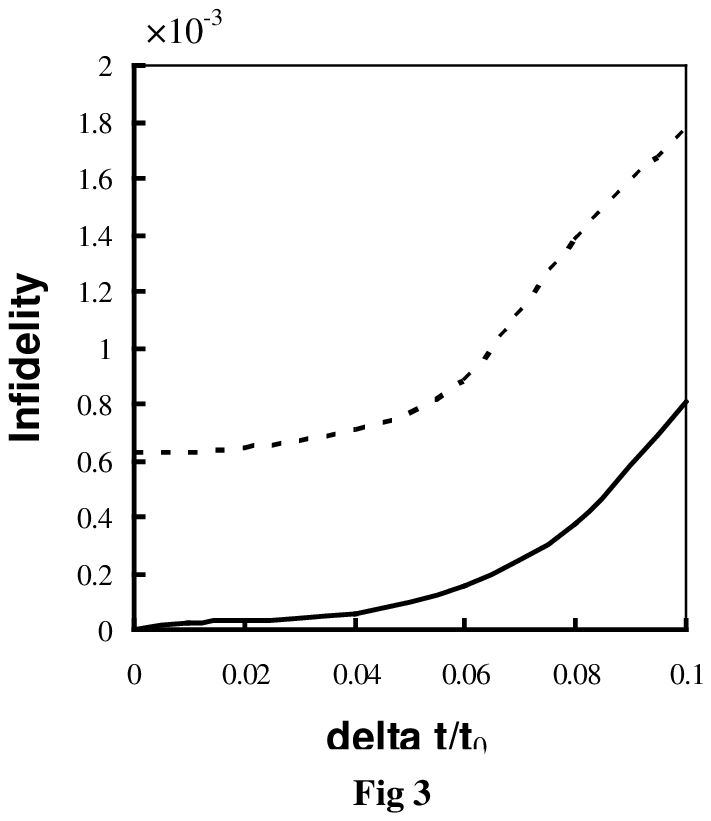}%
\end{center}
\end{figure}

\bigskip%
\begin{figure}
[ptb]
\begin{center}
\includegraphics[
height=11.361in,
width=7.8923in
]%
{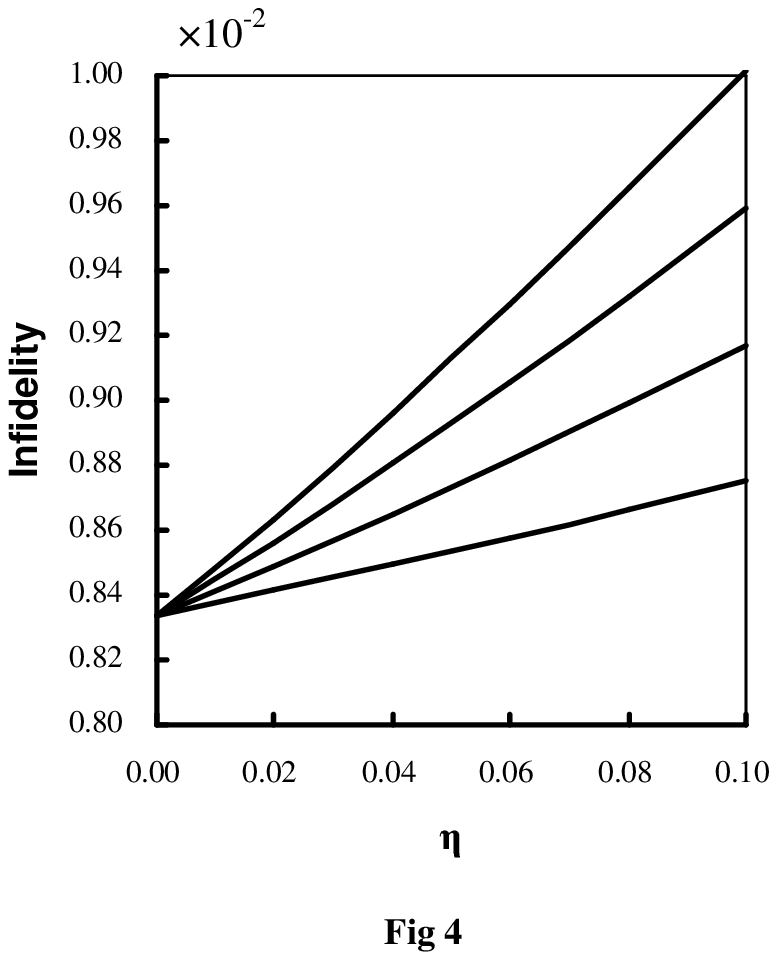}%
\end{center}
\end{figure}

\end{document}